\documentclass[conference]{IEEEtran}
\IEEEoverridecommandlockouts
\usepackage{cite}
\usepackage{amsmath,amssymb,amsfonts}
\usepackage{algorithmic}
\usepackage{graphicx}
\usepackage{textcomp}
\usepackage{xcolor}
\usepackage{booktabs}
\usepackage[normalem]{ulem} 
\usepackage[colorlinks,linkcolor=blue,citecolor=blue]{hyperref}
\usepackage{tikz,xcolor,hyperref}
\def\BibTeX{{\rm B\kern-.05em{\sc i\kern-.025em b}\kern-.08em
    T\kern-.1667em\lower.7ex\hbox{E}\kern-.125emX}}

\DeclareRobustCommand*{\IEEEauthorrefmark}[1]{%
    \raisebox{0pt}[0pt][0pt]{\textsuperscript{\footnotesize\ensuremath{#1}}}}
\begin{document}

\title{Soft-Evidence Fused Graph Neural Network for Cancer Driver Gene Identification across Multi-View Biological Graphs\\

\thanks{This work has been accepted as a regular paper at IEEE International Conference on Bioinformatics and Biomedicine (BIBM) 2025. Source code is available at \url{https://github.com/hanzo2020/SEFGNN}.}
}

\author{
	\IEEEauthorblockN{
		Bang Chen\IEEEauthorrefmark{1}, 
		Lijun Guo{*}\IEEEauthorrefmark{1}, 
		Houli Fan\IEEEauthorrefmark{1}, 
		Wentao He\IEEEauthorrefmark{1} 
		Rong Zhang\IEEEauthorrefmark{1}} 
	\IEEEauthorblockA{\IEEEauthorrefmark{1}The Faculty of Electrical Engineering and Computer Science, Ningbo University, Ningbo, China}
}

\maketitle

\begin{abstract}
Identifying cancer driver genes (CDGs) is essential for understanding cancer mechanisms and developing targeted therapies. Graph neural networks (GNNs) have recently been employed to identify CDGs by capturing patterns in biological interaction networks. However, most GNN-based approaches rely on a single protein-protein interaction (PPI) network, ignoring complementary information from other biological networks. Some studies integrate multiple networks by aligning features with consistency constraints to learn unified gene representations for CDG identification. However, such representation-level fusion often assumes congruent gene relationships across networks, which may overlook network heterogeneity and introduce conflicting information. To address this, we propose Soft-Evidence Fusion Graph Neural Network (SEFGNN), a novel framework for CDG identification across multiple networks at the decision level. Instead of enforcing feature-level consistency, SEFGNN treats each biological network as an independent evidence source and performs uncertainty-aware fusion at the decision level using Dempster-Shafer Theory (DST). To alleviate the risk of overconfidence from DST, we further introduce a Soft Evidence Smoothing (SES) module that improves ranking stability while preserving discriminative performance. Experiments on three cancer datasets show that SEFGNN consistently outperforms state-of-the-art baselines and exhibits strong potential in discovering novel CDGs.
\end{abstract}


\section{Introduction}
Cancer is a highly complex and heterogeneous disease, typically driven by mutations or dysregulation in a small number of key genes known as cancer driver genes (CDGs). Identifying CDGs is essential for understanding the molecular mechanisms of tumorigenesis, developing precise targeted therapies, and improving clinical outcomes~\cite{jang2019transposable}. Traditional experimental techniques like whole-exome sequencing~\cite{luo2021whole} or microarray analysis~\cite{perou2000molecular} have been successfully adopted to screen functional genes, but they suffer from high cost and long processing time, which limit their scalability in large-scale screening tasks.

To enhance efficiency, a wide range of computational methods have been proposed. Early approaches relied on mutation frequency~\cite{dees2012music}, network topology~\cite{bashashati2012drivernet}, or conventional machine learning models~\cite{davoli2013cumulative}, and contributed significantly to CDG discovery. However, these methods often struggle to model the multi-layered, heterogeneous nature of biological data.

With the increasing availability of large-scale omics and biological network data, graph neural networks (GNNs) have emerged as a powerful tool for CDG identification due to their natural capacity to model graph-structured data. Initial GNN-based models, such as EMOGI~\cite{schulte2021integration} and MTGCN~\cite{peng2022improving}, typically integrated multi-omics features into a single biological network using graph convolutional networks (GCNs) to infer candidate driver genes. Later studies further improved GNN architectures or omics feature design. For example, SMG~\cite{cui2023smg} addressed data sparsity via self-supervised masked graph learning, while ECD-CDGI~\cite{wang2024ecd} combined energy-constrained diffusion with attention mechanisms to capture intricate gene relationships. DGGAT~\cite{peng2023integrating} introduced a gating mechanism to capture high-order neighbor interactions and simultaneously detect CDGs and functional gene modules.

Despite these advances, most existing GNN-based methods rely on a single biological network, lacking the ability to integrate multi-source, multi-level, and context-specific biological information. As a result, their capacity to fully capture the complex regulatory mechanisms of cancer remains limited~\cite{chatzianastasis2023explainable}. To address this, recent studies have explored multi-network fusion strategies. For instance, MODIG~\cite{zhao2022modig} constructs a multi-dimensional homogeneous gene network and uses attention mechanisms to aggregate features from various sources. EMGNN~\cite{chatzianastasis2023explainable} and MPIT~\cite{han2023enhancing} align features from different PPI networks to obtain integrated representations. MNGCL~\cite{peng2024multi} employs graph contrastive learning to enforce consistency across networks while learning robust gene representations.

However, existing multi-network approaches often treat different biological networks as homogeneous views and perform feature alignment or attention-based fusion at the representation level. This "forced consistency" strategy implicitly assumes that gene relationships across networks are congruent, which may not hold in practice. For instance, a gene may be labeled as a driver in STRING but lack supporting evidence in PCNet. Such discrepancies reflect distinct biological mechanisms and should be explicitly modeled rather than suppressed. Naively aligning heterogeneous networks or merging their features may obscure unique perspectives or even introduce conflicting signals, thereby compromising the discriminative power and generalizability of the model.

To address these challenges, we propose the Soft-Evidence Fusion Graph Neural Network (SEFGNN), a novel framework for CDG identification across multiple biological networks. Unlike existing multi-network approaches that align node features across views, SEFGNN treats each biological network as a distinct evidence source with heterogeneous structural and semantic properties. Based on Dempster-Shafer Theory (DST), SEFGNN models the prediction from each network as subjective probabilistic evidence, explicitly capturing both belief and uncertainty. This uncertainty-aware fusion strategy enables SEFGNN to integrate multi-view information in a trustworthy manner while preserving the unique contribution of each network. To further alleviate overconfident or polarized outputs often introduced by direct DST fusion, we introduce a Soft Evidence Smoothing (SES) module that reduces output volatility and improves ranking consistency, which is crucial for dependable gene prioritization in practical applications. In summary, our main contributions are as follows:
\begin{enumerate}
\item{We propose SEFGNN, a novel GNN-based framework that performs uncertainty-aware fusion across multiple biological networks, and to the best of our knowledge, is the first to explore CDG identification as a multi-view learning task at the decision-level.}
\item{We design the Soft Evidence Smoothing (SES) module to mitigate extreme outputs introduced by DST-based evidential fusion, enhancing ranking robustness while preserving discriminative accuracy.}
\item{We conduct extensive experiments on three cancer datasets, demonstrating that SEFGNN consistently outperforms state-of-the-art methods and exhibits promising potential in discovering novel CDGs.}
\end{enumerate}

\section{MATERIALS AND METHODS}
\subsection{Problem Statement}
Let \( G = \{G^{(1)}, G^{(2)}, \ldots, G^{(N)}\} \) denote a set of \(N\) biological networks, where each graph \( G^{(i)} = (V, E_i, X) \) shares the same node set \( V = \{v_1, \ldots, v_{|V|}\} \) and feature matrix \( X \in \mathbb{R}^{|V| \times m} \), but differs in its edge set \( E_i \subseteq V \times V \), reflecting the heterogeneity across biological conditions. Each node represents a gene, and each row of \( X \) encodes an \(m\)-dimensional multi-omics feature vector $\mathbf{x}_i \in \mathbb{R}^m$.

Given the multi-network input \( G \), cancer driver gene identification is formulated as a node-level binary classification task. The goal is to learn a function \( f: G \to \{0,1\}^{|V|} \) that predicts whether each gene \( v_i \in V \) is a cancer driver gene.

\subsection{Datasets and Processing}
In this study, we adopt the dataset from \cite{han2023enhancing}, which includes 16,165 protein-coding genes, each annotated with ten features derived from six omics types: ATAC-seq (1 feature), CTCF (3), H3K4me3 (2), H3K27ac (2), CNV (1), and SNV (1). The dataset also provides five protein–protein interaction (PPI) networks, each offering a distinct topological view of gene interactions, with edge counts as follows: CPDB (273,765), STRING (253,535), PCNet (2,192,197), iRefIndex (342,006), and Multinet (83,766).
Following the labeling strategy in \cite{schulte2021integration} and \cite{han2023enhancing}, positive samples include genes from NCG \cite{repana2019network}, CGC \cite{sondka2018cosmic}, DigSee \cite{kim2013digsee}, DisGeNet \cite{pinero2016disgenet}, and DriverDBv4 \cite{cheng2014driverdb}. Negative genes are defined via recursive exclusion: removing genes in NCG, CGC, OMIM, and those predicted as cancer-related by MSigDB. The resulting MCF7, K562, and A549 datasets contain 379/1581, 610/1838, and 425/2557 positive/negative genes, respectively, and are randomly split into training, validation, and test sets (6:2:2).

\subsection{Model Architecture and Components}
\begin{figure*}[htbp]
\centerline{\includegraphics[scale=.28]{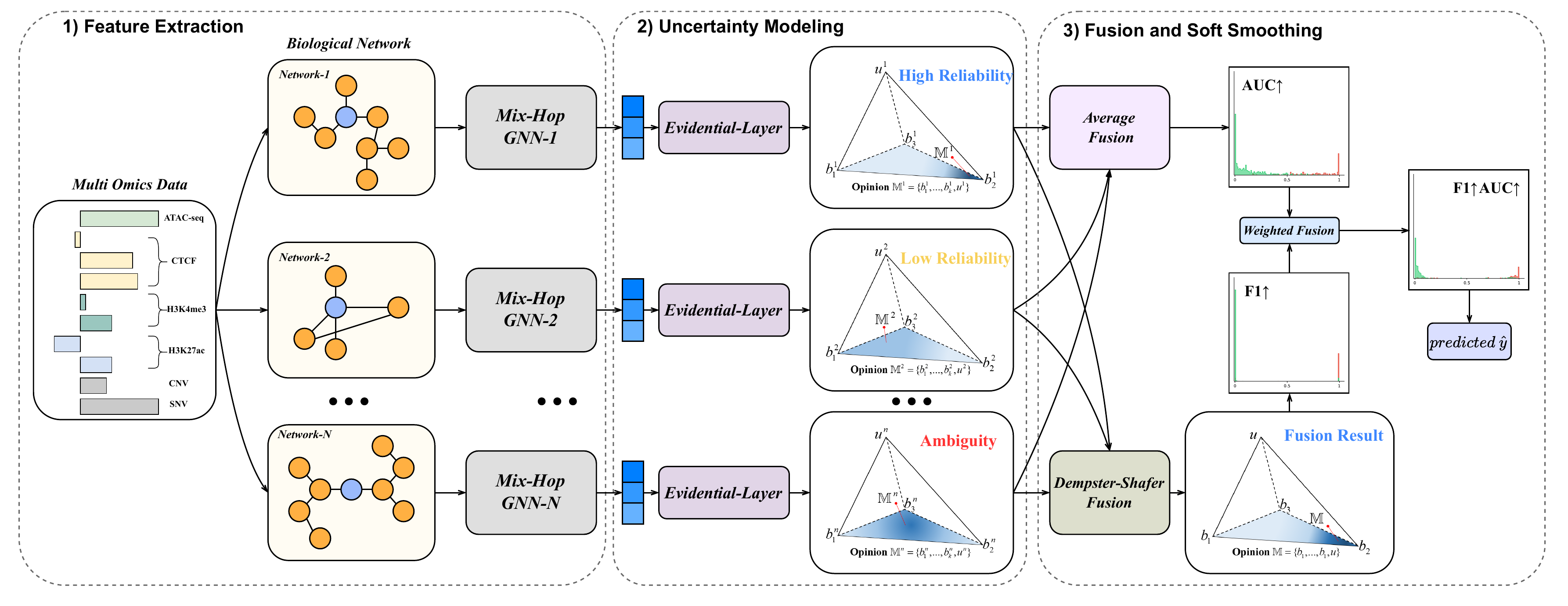}}
\caption{Overview of the SEFGNN architecture. The model extracts features from multiple biological networks using independent GNNs, models uncertainty via evidential layers, and fuses predictions through Dempster-Shafer theory with a soft evidence smoothing module.}
\label{fig1}
\end{figure*}
As illustrated in Fig.~\ref{fig1}, the overall architecture of our model consists of three main components. First, a set of independent GNNs is employed to extract node representations from multiple biological network views. Then, a view-specific evidential neural network transforms these representations into Dirichlet-distributed evidence, enabling uncertainty modeling through subjective logic. Finally, the evidence from different networks is integrated via Dempster-Shafer theory, and a soft evidence smoothing module is applied to produce the final prediction.

\subsubsection{MixHop-Based Feature Extraction Module}

We first employ a MixHop-based feature extraction module \cite{abu2019mixhop} to learn gene representations from $N$ biological networks. For each gene node $v_i$, the initial input is a $m$-dimensional multi-omics feature vector $\mathbf{x}_i \in \mathbb{R}^m$. To capture information from multi-hop neighborhoods, MixHop convolutions aggregate features from neighbors at various hop distances, with each hop-specific aggregation weighted and concatenated to produce a richer representation.

Specifically, assuming that the MixHop convolution aggregates information from a set of hop distances $\mathcal{P}$, the message propagation and aggregation at each layer can be formulated as:

\begin{equation}
\mathbf{X}' = \mathop{\|}_{p \in \mathcal{P}} \left( \hat{\mathbf{D}}^{-\frac{1}{2}} \hat{\mathbf{A}} \hat{\mathbf{D}}^{-\frac{1}{2}} \right)^p \mathbf{X} \boldsymbol{\Theta}_p\,,
\label{eq:mixhop}
\end{equation}
Here, $\hat{\mathbf{A}} = \mathbf{A} + \mathbf{I}$ denotes the adjacency matrix with added self-loops, and $\hat{\mathbf{D}}$ is the corresponding degree matrix. The input feature matrix is $\mathbf{X} \in \mathbb{R}^{|V| \times m}$, and $\boldsymbol{\Theta}_p$ denotes the learnable transformation parameters for the $p$-hop neighborhood. The operator $\|$ represents feature concatenation across hops.

Through this process, node features from neighborhoods of different hop distances are effectively aggregated and fused, enhancing the discrimination power of node representations.

Finally, for each gene $v$, we obtain a set of node representations from $N$ distinct biological network views: $\{\mathbf{z}^{(1)}, \mathbf{z}^{(2)},\ldots, \mathbf{z}^{(N)}\}$, which are subsequently used as inputs for uncertainty modeling and multi-view reasoning.

\subsubsection{Uncertainty Modeling and Multi-View Fusion Based on Evidential Theory}

To effectively integrate predictions from multi-view biological networks, we adopt a Dirichlet-based evidential learning framework inspired by \cite{han2022trusted}. Specifically, we model each view's output as a subjective opinion over class labels using the Dirichlet distribution and subjective logic.

Given the feature representation $\mathbf{z}^{(n)}$ of a gene $v$ extracted from the $n$-th biological network, where $n \in \{1, 2, \ldots, N\}$, an evidence network composed of a fully connected layer followed by a \textit{Softplus} activation is used to produce a non-negative evidence vector $\mathbf{e}^{(n)} = [e^{(n)}_1, \ldots, e^{(n)}_K]$, where $K$ is the number of classes (e.g., $K=2$ in CDG identification). The parameters of the Dirichlet distribution are then given by:
\begin{equation}
\boldsymbol{\alpha}^{(n)} = \mathbf{e}^{(n)} + 1\,.
\end{equation}

Following the principles of subjective logic \cite{josang2016subjective}, the Dirichlet parameters $\boldsymbol{\alpha}^{(n)}$ can be further transformed into an opinion representation:
\begin{equation}
\mathbb{M}^{(n)} = \{ b_1^{(n)}, \ldots, b_K^{(n)}, u^{(n)} \}\,,
\end{equation}
where each belief mass and the associated uncertainty are computed as:
\begin{equation}
b_k^{(n)} = \frac{\alpha_k^{(n)} - 1}{S^{(n)}} = \frac{e_k^{(n)}}{S^{(n)}}, \quad
u^{(n)} = \frac{K}{S^{(n)}}\,.
\label{eq:belief_uncertainty}
\end{equation}
Here, $S^{(n)} = \sum_{k=1}^{K} \alpha_k^{(n)}$ denotes the Dirichlet strength. The term $u^{(n)} \geq 0$ quantifies the uncertainty of the $n$-th view, while $b_k^{(n)} \geq 0$ reflects the belief mass assigned to class $k$.

Subsequently, following \cite{han2022trusted}, we integrate the opinions from different biological networks based on the Dempster-Shafer (DS) evidence fusion rule \cite{josang2012interpretation}. For instance, given two subjective opinions $\mathbb{M}^{(1)} = \left\{ \{b_{k}^{(1)}\}_{k=1}^{K},\ u^{(1)} \right\}$ and $\mathbb{M}^{(2)} = \left\{ \{b_{k}^{(2)}\}_{k=1}^{K},\ u^{(2)} \right\}$, the DS theory fuses them into a combined opinion $\mathbb{M} = \left\{ \{b_{k}\}_{k=1}^{K},\ u \right\}$, with the following calculation:
\begin{equation}
b_{k} = \frac{1}{C} \left( b_{k}^{(1)} b_{k}^{(2)} + b_{k}^{(1)} u^{(2)} + b_{k}^{(2)} u^{(1)} \right), \quad
u = \frac{1}{C} u^{(1)} u^{(2)}\,,
\end{equation}
where $C = 1 - \sum\nolimits_{i \ne j} b_{i}^{(1)} b_{j}^{(2)}$ is the normalization factor.

This fusion rule possesses the following key properties:  
1) When both views exhibit high uncertainty (i.e., $u^{(1)}$ and $u^{(2)}$ are large), the fused result also has high uncertainty (i.e., $b_{k}$ is small), and vice versa.  
2) If there is a large difference in uncertainty between the two, the result tends to favor the more confident (i.e., less uncertain) view.  
3) In the case of conflicting beliefs, both $C$ and $u$ increase accordingly.

In practice, we recursively apply pairwise fusion to all opinions from the biological networks. The final global fused opinion is converted into Dirichlet parameters $\boldsymbol{\alpha}^{\text{DS}} = [\alpha_1^{\text{DS}}, \ldots, \alpha_K^{\text{DS}}]$ as follows:
\begin{equation}
S = \frac{K}{u}, \quad
e_k = b_k \cdot S, \quad
\alpha_k^{\text{DS}} = e_k + 1\,.
\end{equation}

Finally, we apply a softmax function and take the probability corresponding to the positive class to obtain the final fused prediction:
\begin{equation}
y_{\text{DS}} = \text{softmax}(\boldsymbol{\alpha}^{\text{DS}})_{[1]}\,.
\end{equation}

\subsubsection{Soft Evidence Smoothing (SES) Module and Final Output}
\begin{figure}[htbp]
\centerline{\includegraphics[scale=.5]{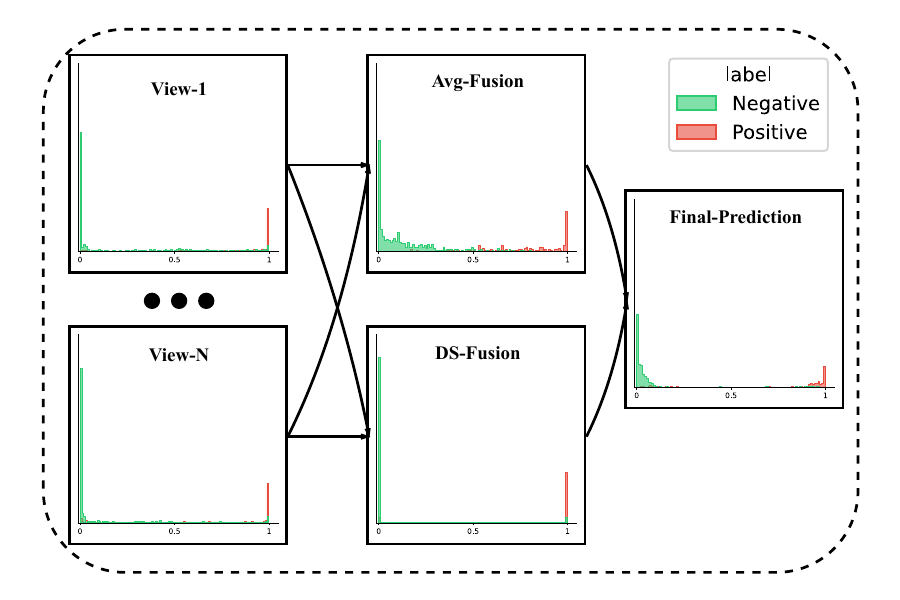}}
\caption{Prediction distributions for Avg-Fusion and DS-Fusion. DS-Fusion increases polarization, improving accuracy but harming ranking.}
\label{fig2}
\end{figure}
Although the Dempster-Shafer (DS) fusion strategy effectively integrates opinions from multiple biological networks and offers advantages in classification accuracy, we observe that its output distribution tends to be overly polarized in the CDG identification task. This phenomenon adversely affects the model's stability in ranking-oriented evaluations such as AUC.

Figure~\ref{fig2} illustrates five representative prediction distributions. In each subfigure, green bars denote samples that are truly non-driver genes, and red bars denote true driver genes. The left two subfigures show the prediction distributions of two individual views (View-1 and View-N), both of which suffer from limited performance and noticeable classification errors. The top middle subfigure presents the result of average fusion (Avg-Fusion), where the class probabilities from multiple views are averaged. Although this approach exhibits lower classification accuracy, the resulting probability distribution is more continuous and retains better ranking capability. In contrast, the bottom middle subfigure shows the prediction distribution after DS fusion (DS-Fusion), which is heavily concentrated near 0 and 1. This overly polarized distribution leads to a noticeable drop in ranking performance, particularly in AUC. We speculate that such degradation occurs because when an individual view produces high-confidence mispredictions, its influence on the fusion process is disproportionately amplified, thereby compromising the overall ranking stability.

To address this issue, we propose a Soft Evidence Smoothing (SES) module that mitigates the extreme output tendencies. Specifically, we first compute the probability that a given gene is a positive sample by applying a softmax operation to the Dirichlet parameters $\boldsymbol{\alpha}^{(n)}$ from each view $(n)$ and taking the value at index 1:
\begin{equation}
y_{\text{avg}} = \frac{1}{N} \sum\limits_{n=1}^{N} \text{softmax}(\boldsymbol{\alpha}^{(n)})_{[1]}\,.
\end{equation}

Finally, the model output is obtained by a weighted combination of the DS-fused prediction and the average prediction:
\begin{equation}
\widehat{y} = \gamma \cdot y_{\text{DS}} + (1 - \gamma) \cdot y_{\text{avg}}\,,
\end{equation}
where $\gamma \in (0, 1)$ is a learnable parameter. As shown in the rightmost subfigure of Figure~\textit{x}, the smoothed output not only preserves classification performance but also significantly improves ranking performance (AUC).

\subsubsection{Loss Function}

The proposed model is trained in an end-to-end manner. For each biological network view $(n)$, we adopt the Integrated Cross-Entropy (ICE) loss to supervise the estimated Dirichlet parameters. Specifically, the ICE loss is defined as:
\begin{equation}
\begin{split}
\mathcal{L}_{\mathrm{ice}}^{(n)} =\ & \mathbb{E}_{\mathbf{p}^{(n)} \sim \mathrm{Dir}(\mathbf{p}^{(n)} \mid \boldsymbol{\alpha}^{(n)})} 
\left[\mathcal{L}_{\mathrm{CE}}(\mathbf{p}^{(n)}, \mathbf{y}^{(n)})\right] \\
=\ & \sum_{k=1}^{K} y_k^{(n)} \left( \psi(S^{(n)}) - \psi(\alpha_k^{(n)}) \right)\,,
\end{split}
\end{equation}
where $\psi(\cdot)$ denotes the Digamma function, $\mathbf{y}^{(n)}$ is a one-hot encoded ground-truth label vector, and $S^{(n)} = \sum_{k=1}^{K} \alpha_k^{(n)}$ is the concentration parameter of the Dirichlet distribution for view $(n)$.

In addition, a Kullback-Leibler (KL) divergence term is introduced as a regularizer to model uncertainty. The complete loss for view $(n)$ is defined as:
\begin{equation}
\mathcal{L}^{(n)} = \mathcal{L}_{\mathrm{ice}}^{(n)} + \lambda \cdot \mathrm{KL} \left[ \mathrm{Dir}(\mathbf{p}^{(n)}; \widetilde{\boldsymbol{\alpha}}^{(n)}) \parallel \mathrm{Dir}(\mathbf{p}^{(n)} \mid \mathbf{1}) \right]\,,
\end{equation}
where $\lambda$ is a gradually increasing balancing coefficient during training, and $\widetilde{\boldsymbol{\alpha}}^{(n)} = \mathbf{y}^{(n)} + (1 - \mathbf{y}^{(n)}) \odot \boldsymbol{\alpha}^{(n)}$ represents a target Dirichlet parameter adjusted by the ground-truth label, with $\odot$ denoting element-wise multiplication.

To ensure that both individual biological networks and the DS-fused output are effectively optimized, we compute losses for each view as well as for the fusion output. Furthermore, we introduce an entropy-based regularization term to constrain the soft evidence weighting coefficient $\gamma$, preventing it from collapsing to extreme values (0 or 1) and thereby promoting balanced fusion:
\begin{equation}
\mathcal{L}_{\mathrm{entropy}} = -\left[ \gamma \log(\gamma + \varepsilon) + (1 - \gamma) \log(1 - \gamma + \varepsilon) \right]\,,
\end{equation}
where $\varepsilon$ is a small constant for numerical stability.

The overall training objective of the model is given by:
\begin{equation}
\mathcal{L}_{\mathrm{Overall}} = \mathcal{L}^{\mathrm{DS}} + \sum_{n=1}^{N} \mathcal{L}^{(n)} + \lambda_1 \mathcal{L}_{\mathrm{entropy}} + \lambda_2 \mathcal{L}_{\mathrm{len}} + \lambda_3 \mathcal{L}_{\Theta}\,,
\end{equation}
Here, $\mathcal{L}^{\mathrm{DS}}$ denotes the loss of the DS-fused output. $\mathcal{L}_{\mathrm{len}} = \sum_{i=1}^{N} \| Z_{i} \|_{F}$ and $\mathcal{L}_{\Theta} = \sum_{i} \| \Theta_{i} \|_{F}^{2}$ are regularization terms applied respectively to the output features $Z_{i}$ of the GNNs and all trainable parameters $\Theta_{i}$ in the model, aiming to enhance training stability. The coefficients $\lambda_1$, $\lambda_2$, and $\lambda_3$ are tunable hyperparameters used to balance the influence of these auxiliary losses.

\section{RESULTS AND ANALYSIS}

\subsection{Baselines and Parameters setting}
To comprehensively evaluate the performance of our proposed model, we compared it against eight representative baseline methods, categorized into three groups: (1) traditional GNNs, including GCN \cite{kipf2017semisupervised}, GTN \cite{yun2019graph}, and GAT \cite{veličković2018graph}; (2) single-network methods, including EMOGI \cite{schulte2021integration}, MTGCN \cite{peng2022improving}, and ECD-CDGI \cite{wang2024ecd}; and (3) multi-network approaches, including MNGCL \cite{peng2024multi} and MPIT \cite{han2023enhancing}.

All models were trained for 300 epochs using the Adam optimizer with a batch size of 128 and a learning rate of $1 \times 10^{-4}$. Early stopping was applied based on validation performance. For our model, we set $\lambda_1 = \lambda_2 = \lambda_3 = 1 \times 10^{-5}$. The hidden dimension of gene embeddings was fixed at 48 across all GNN-based methods. GCN, GAT, and GTN were implemented using PyTorch Geometric, while the other baselines were reproduced from their official repositories with default or recommended settings. Finally, all experiments were conducted with three different random seeds, and the average performance of each model was reported.
\begin{table*}[!t]
\caption{Performance comparison across three cancer datasets. \underline{Underlined} values denote the best results among all baselines. \uwave{Wavy underlines} indicate the best performance excluding SEFGNN. \pmb{Bold} values highlight the overall best performance.}
\begin{center}
\begin{tabular}{ccccccccccccc}
\toprule
 & \multicolumn{4}{c}{\pmb{Breast Cancer}} & \multicolumn{4}{c}{\pmb{Leukemia}} & \multicolumn{4}{c}{\pmb{Lung Cancer}}\\
\cmidrule{2-13} 
& \pmb{F1} & \pmb{ACC}& \pmb{AUC} & \pmb{AUPRC} & \pmb{F1} & \pmb{ACC}& \pmb{AUC} & \pmb{AUPRC} & \pmb{F1} & \pmb{ACC} & \pmb{AUC} & \pmb{AUPRC}\\
\midrule
GCN & 0.7768 & 0.9184 & 0.9436 & 0.8494 & 0.7989 & 0.8979 & 0.9377 & 0.8571 & 0.6725 & 0.9190 & 0.9163 & 0.7468\\
GAT & 0.6344 & 0.8070 & 0.8903 & 0.5433 & 0.7374 & 0.8483 & 0.9161 & 0.7554 & 0.5147 & 0.6689 & 0.7095 & 0.4745\\
GTN & 0.7907 & 0.9099 & 0.9627 & 0.8888 & 0.7714 & 0.8544 & \uwave{\underline{0.9759}} & \uwave{\underline{0.9470}} & 0.7090 & 0.9056 & 0.9381 & 0.7784\\
\midrule
EMOGI & 0.8230 & 0.9294 & 0.9672 & 0.8798 & 0.8411 & 0.9156 & 0.9744 & 0.9455 & 0.7399 & 0.9246 & 0.9347 & 0.7296\\
ECD-CDGI & 0.8547 & 0.9439 & \uwave{\underline{0.9786}} & 0.9406 & 0.8514 & 0.9245 & 0.9694 & 0.9326 & \underline{0.7718} & 0.9403 & \uwave{\underline{0.9541}} & \uwave{\underline{0.8552}}\\
MTGCN & 0.7897 & 0.9175 & 0.9590 & 0.8828 & 0.7523 & 0.8524 & 0.9478 & 0.8846 & 0.7143 & 0.9246 & 0.9349 & 0.7981\\
\midrule
MNGCL & 0.8242 & 0.9354 & 0.9670 & 0.9036 & \underline{0.8613} & \underline{0.9299} & 0.9672 & 0.9261 & 0.7699 & \underline{0.9419} & 0.9327 & 0.8335\\
MPIT & \underline{0.8681} & \underline{0.9473} & 0.9720 & \uwave{\underline{0.9442}} & 0.8414 & 0.9156 & 0.9682 & 0.9132 & 0.7448 & 0.9235 & 0.9419 & 0.8205\\
\midrule
EFGNN & \uwave{0.8904} & \uwave{0.9600} & 0.9501 & 0.9059 &  \uwave{0.8780} & \uwave{0.9388} & 0.9659 & 0.8943 & \uwave{0.8129} & \pmb{0.9525} & 0.9220 & 0.8489\\
SEFGNN & \pmb{0.8915} & \pmb{0.9609} & \pmb{0.9856} & \pmb{0.9604} & \pmb{0.8859} & \pmb{0.9422} & \pmb{0.9788} & \pmb{0.9504} & \pmb{0.8158} & \pmb{0.9525} & \pmb{0.9589} & \pmb{0.8950}\\
\midrule
\midrule
improve1 & 2.70\% & 1.44\% & 0.71\% & 1.72\% & 2.86\% & 1.32\% & 0.30\% & 0.36\% & 5.70\% & 1.13\% & 0.50\% & 5.02\%\\
improve2 & 0.12\% & 0.09\% & 3.74\% & 6.02\% & 0.90\% & 0.36\% & 1.34\% & 6.27\% & 0.36\% & 0.00\% & 4.00\% & 5.43\%\\
\bottomrule
\end{tabular}
\label{tab1}
\end{center}
\end{table*}
\subsection{Performance evaluation}
Table~\ref{tab1} summarizes the performance of all methods on three cancer types. EFGNN is an ablation variant of SEFGNN without the Soft Evidence Smoothing (SES) module. Overall, SEFGNN consistently outperforms all existing approaches across datasets and evaluation metrics. Among the baselines, single-network methods generally outperform traditional GNNs such as GCN and GAT, possibly due to task-specific architectural enhancements. Furthermore, multi-network integration methods such as MNGCL and MPIT achieve even better performance, highlighting the effectiveness of integrating multiple biological networks. Specifically, improve1 represents the performance improvement of SEFGNN over the previous best, while improve2 reflects the additional gain over EFGNN, brought by the proposed SES module. Notably, although EFGNN significantly improves F1 scores compared to other baselines, its ranking performance (AUC, AUPRC) deteriorates, suggesting overconfident predictions under uncertainty. By smoothing overly confident evidence, the SES module in SEFGNN improves the robustness of ranking-based metrics while maintaining high classification accuracy.

\subsection{ablation Study}
\begin{table}[!ht]
\caption{Ablation study demonstrating the advantage of trusted fusion over single-network settings.}
\centering
\begin{tabular}{ccccccc}
\toprule
 & \multicolumn{2}{c}{\pmb{Breast Cancer}} & \multicolumn{2}{c}{\pmb{Leukemia}} & \multicolumn{2}{c}{\pmb{Lung Cancer}}\\
\cmidrule{2-7} 
& \pmb{F1} & \pmb{AUC} & \pmb{F1} & \pmb{AUC} & \pmb{F1} & \pmb{AUC} \\
\midrule
CPDB & 0.7669  & 0.9552  & 0.7489  & 0.9294  & 0.7073  & 0.9127 \\
STRING & 0.7600  & 0.9442  & 0.6916  & 0.9297  & 0.5410  & 0.8979 \\
PCNet & 0.7771  & 0.9157  & 0.7935  & 0.9225  & 0.6108  & 0.8930 \\
iRefIndex & 0.8516  & 0.9676  & 0.8614  & 0.9741  & 0.6759  & 0.9461 \\
Multinet & 0.8050  & 0.9507  & 0.7698  & 0.9254  & 0.6744  & 0.9001 \\
EFGNN & 0.8904  & 0.9501  & 0.8780  & 0.9659  & 0.8129  & 0.9220 \\
SEFGNN & 0.8915  & 0.9856  & 0.8859  & 0.9788  & 0.8158  & 0.9589 \\
\bottomrule
\end{tabular}
\label{tab2}
\end{table}
Table~\ref{tab2} compares SEFGNN with its single-network variants and ablated versions. The first five rows show the results using individual biological networks, all of which perform worse than EFGNN and SEFGNN, highlighting the benefit of integrating complementary information. EFGNN, which fuses network predictions without SES, consistently outperforms single-network models. SEFGNN further improves upon EFGNN, especially in AUC, by smoothing overconfident outputs, leading to more stable and accurate predictions. These results confirm the effectiveness of our uncertainty-aware fusion strategy in CDG identification.

\subsection{Identifying New Cancer Genes}
Given that the currently known set of CDGs is still incomplete, we further investigate the capability of SEFGNN to discover novel cancer drivers. Specifically, we train the model using all confirmed positive genes (CDGs) and negative genes (non-CDGs) in each of the three cancer datasets, and then apply it to predict the remaining unlabeled genes, aiming to identify potential cancer drivers among these unannotated candidates.

\begin{table}[!ht]
\centering
\small
\caption{Co-citation validation of predicted genes. CG: CCGD; CM: CancerMine; CA: CancerAlterome.}
\resizebox{\columnwidth}{!}{
\begin{tabular}{cccccccccccccccc}
\toprule
 &\multicolumn{5}{c}{\textbf{MCF7}} &\multicolumn{5}{c}{\textbf{K562}} &\multicolumn{5}{c}{\textbf{A549}}\\
\cmidrule(lr){2-6}\cmidrule(lr){7-11}\cmidrule(lr){12-16}
rank & gene & CG & CM & CA & Breast & gene & CG & CM & CA & Leukemia & gene & CG & CM & CA & Lung \\
\midrule
1  & KDM1A & 12 & 24 & 27 & 4 & HSPG2 & 1 & 4 & 9 & 2 & CAV1 & 1 & 38 & 44 & 14 \\
2  & SFN   & 1 & 14 & 8 & 4 & CDC42 & 9 & 20 & 15 & 5 & CDC42 & 9 & 20 & 1 & 3 \\
3  & YAP1  & 7 & 458 & 137 & 75 & UBC & 2 & 3 & 2 & 1 & HDAC1 & 0 & 17 & 27 & 8 \\
4  & PTPN11& 3 & 106 & 55 & 14 & SMAD4 & 17 & 231 & 125 & 3 & JUN & 1 & 251 & 31 & 28 \\
5  & PSMD13& 2 & 0 & 0 & 0 & STK11 & 6 & 230 & 116 & 2 & CDK1 & 0 & 194 & 28 & 29 \\
6  & COL4A1& 1 & 2 & 5 & 2 & TRAF1 & 1 & 0 & 1 & 1 & KDM1A & 12 & 24 & 27 & 2 \\
7  & HSP90AA1 & 0 & 19 & 42 & 4 & CUL2 & 3 & 2 & 0 & 1 & GSK3B & 28 & 3 & 66 & 5 \\
8  & STAT1 & 1 & 57 & 40 & 14 & SMAD2 & 8 & 14 & 16 & 2 & CUL1 & 16 & 1 & 5 & 0 \\
9  & PPP1CC& 4 & 0 & 0 & 0 & NGFR & 0 & 30 & 6 & 1 & RPA2 & 1 & 0 & 1 & 0 \\
10 & ARRB2 & 0 & 0 & 2 & 0 & KPNB1 & 9 & 2 & 4 & 1 & TBP & 2 & 60 & 2 & 5 \\
\bottomrule
\end{tabular}
}
\label{tab3}
\end{table}

\subsubsection{Co-citation Analysis}
Following the methodology in \cite{li2025towards}, we conducted a co-citation analysis of the top 10 high-confidence genes predicted by SEFGNN in three cell lines: MCF7 (breast), K562 (leukemia), and A549 (lung). The results are summarized in Table~\ref{tab3}.

Predicted genes were cross-validated using three cancer-related databases: Cancer Gene Census (CG), CancerMine (CM), and CancerAlterome (CA). Columns labeled \textbf{CG}, \textbf{CM}, and \textbf{CA} denote the number of co-cited publications, while \textbf{Breast}, \textbf{Leukemia}, and \textbf{Lung} indicate gene–cancer co-occurrence frequencies in the literature.

The statistical results illustrate that most of the predicted genes have varying degrees of literature support in at least one of the databases. Specifically, 27 genes are recorded in CG, 29 in CM, and 26 in CA. The gene-cancer type co-occurrence analysis also demonstrates strong correspondence: for the MCF7 model, 7 genes co-occurred with the keyword ``Breast cancer''; for the K562 model, all top-ranked genes were found to co-occur with ``Leukemia''; and for the A549 model, 8 genes co-occurred with ``Lung cancer'' in the literature.

Notably, some genes such as PSMD13 and RPA2 exhibit limited database evidence yet are functionally linked to key biological processes including proteasome regulation and DNA repair \cite{li2025comprehensive, chen2017upregulation}. These results suggest that SEFGNN can identify biologically relevant but underreported cancer driver genes.

\subsubsection{Drug sensitivity analysis}

Given that drug sensitivity can reveal the potential roles of CDGs in regulating therapeutic response, we further investigated the associations between the top-ranked genes predicted by SEFGNN and anticancer drug sensitivity, following the approach described in \cite{li2025towards}. Specifically, we employed the drug sensitivity analysis module of the Gene Set Cancer Analysis (GSCA) platform \cite{liu2023gsca} to examine the correlations between gene expression and drug sensitivity. For each dataset (MCF7, K562, and A549), we selected the top 10 candidate genes ranked by SEFGNN and assessed their associations with multiple anticancer agents.

\begin{figure}[htbp] \centerline{\includegraphics[scale=.33]{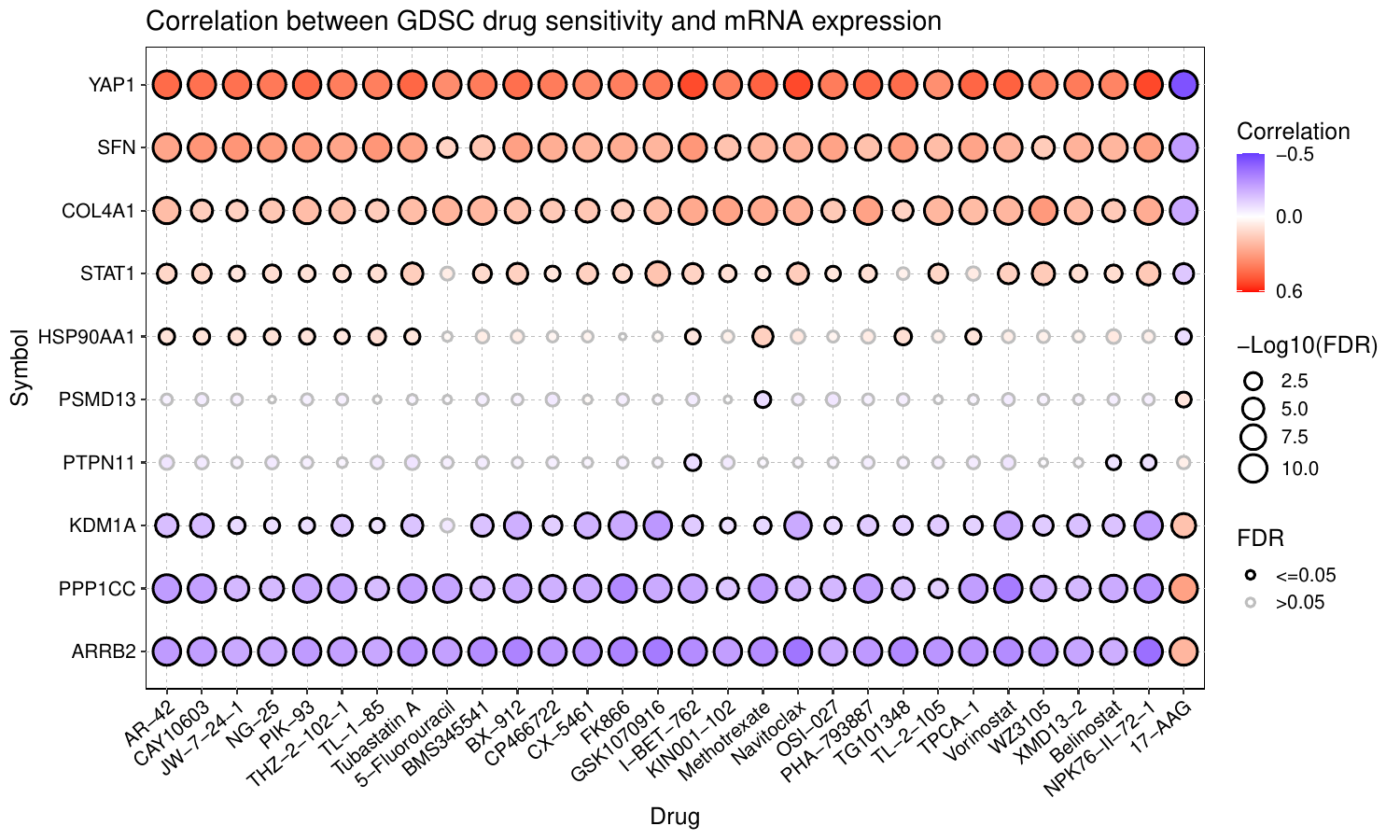}} \caption{Correlation between drug sensitivity and mRNA expression for the top 10 predicted genes in MCF7 dataset.} \label{fig3} \end{figure} \begin{figure}[htbp] \centerline{\includegraphics[scale=.33]{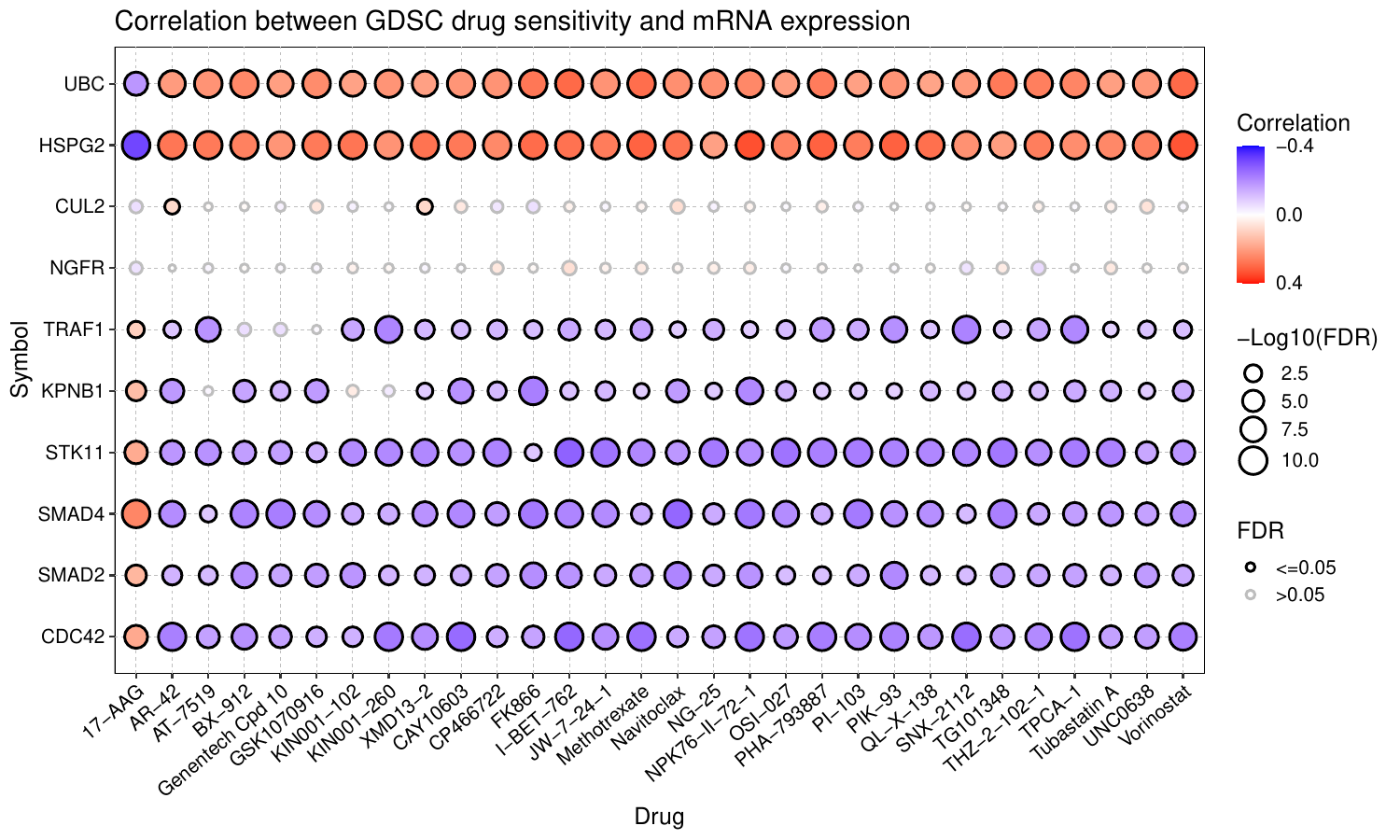}} \caption{Correlation between drug sensitivity and mRNA expression for the top 10 predicted genes in K562 dataset.} \label{fig4} \end{figure} \begin{figure}[htbp] \centerline{\includegraphics[scale=.33]{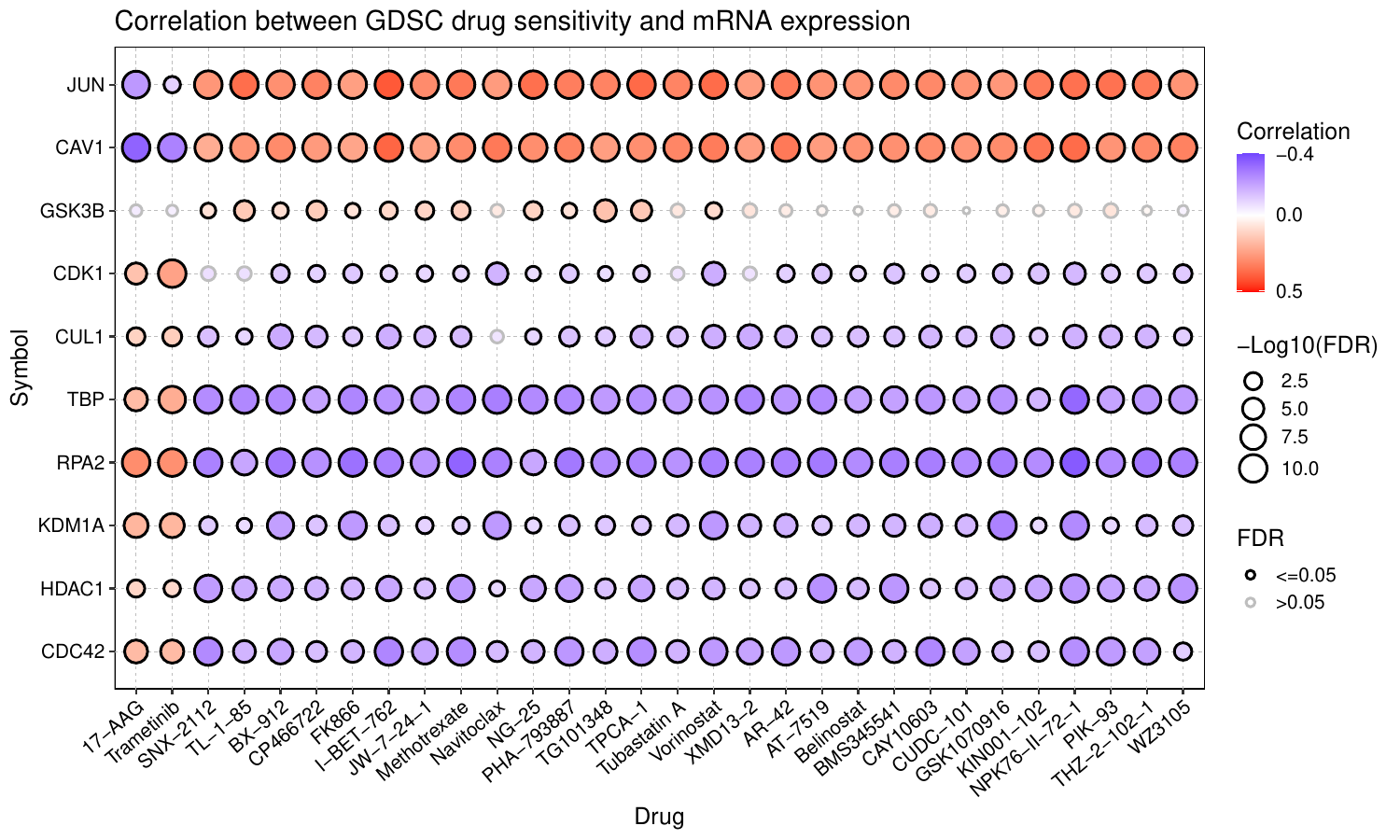}} \caption{Correlation between drug sensitivity and mRNA expression for the top 10 predicted genes in A549 dataset.} \label{fig5} \end{figure}

As shown in Fig. \ref{fig3} - \ref{fig5}, the results are displayed as bubble plots, where color indicates correlation direction and strength (red for positive, blue for negative), bubble size reflects statistical significance ($-\log_{10}(\mathrm{FDR})$), and border thickness denotes FDR thresholds (solid for $\leq$ 0.05). In the MCF7 dataset, several genes show significant correlations with a variety of anticancer drugs. For instance, the expression level of YAP1 is positively correlated with multiple compounds. Among them, Belinostat is a histone deacetylase inhibitor that has demonstrated therapeutic efficacy in breast cancer by altering the epigenetic landscape of cancer cells, thereby affecting gene transcription and cell cycle progression. Another correlated drug, Navitoclax, functions by inhibiting proteins of the BCL-2 family, which play a central role in the survival of malignant cells, particularly in hormone-responsive breast cancers.

In the K562 and A549 datasets, most of the top-ranked genes are also significantly associated with drug sensitivity. Among the drugs showing strong correlations with these genes, representative examples include UNC0638 and AR-42. UNC0638 is a histone deacetylase inhibitor that regulates the acetylation of both histone and non-histone proteins, leading to changes in gene expression and reduced proliferation in leukemia and lung cancer cells. AR-42, a phenylbutyrate-based compound with both anti-inflammatory and antitumor properties, modulates multiple signaling pathways related to cell survival and apoptosis, and has demonstrated broad-spectrum anticancer activity. Overall, these findings further support the potential of our model to uncover biologically relevant CDGs.

\section{Conclusion}
This paper presents SEFGNN, a novel GNN framework for CDG identification across multiple heterogeneous biological networks. Instead of enforcing feature-level alignment, SEFGNN treats each PPI network as an independent evidence source and performs uncertainty-aware fusion at the decision level using Dempster-Shafer Theory (DST). To address the issue of polarized outputs in DST, we design a Soft Evidence Smoothing (SES) module that improves ranking stability without compromising classification performance. Extensive experiments on three cancer types validate the effectiveness and generalizability of our approach.

Future directions include incorporating more diverse biological networks, leveraging uncertainty to enhance model interpretability, and introducing pseudo-views into the fusion process. Furthermore, enhancing the DST-based evidence fusion mechanism itself could help better resolve conflicts among network predictions and improve the robustness of multi-network integration.
\section*{Acknowledgment}
The authors would like to respect and thank all reviewers for their constructive and helpful review.

\bibliographystyle{IEEEtran}
\bibliography{IEEEabrv,references}

\end{document}